\documentstyle[prl,aps]{revtex}
\begin{document}
\draft
\begin{titlepage} 
\title{Hydrodynamic Collimation of GRB Fireballs}
\author{Amir Levinson}
\address{ School of Physics and Astronomy, Tel Aviv University, 
Tel Aviv 69978, Israel}
\author{David Eichler}
\address{Physics Department, Ben-Gurion University, Beer-Sheva}
\maketitle
\begin{abstract}
Analytic solutions are presented for the hydrodynamic collimation
of a relativistic fireball by a surrounding baryonic wind
emanating from a torus. The opening angle is shown to be
the ratio of the  power output of the inner fireball to that of
the exterior baryonic wind. The gamma ray burst 990123 might thus
be interpreted as a baryon-pure jet with an energy output  of
order  10$^{50}$ erg or less, collimated by a baryonic wind from
a torus with an energy output of order $10^{52}$ erg.
\end{abstract}
\end{titlepage}

Recent observations of GRB's and their afterglows appear to
indicate that GRB ejecta are collimated.  In particular, i) the
high fluences exhibited by several bursts, notably GRB 990123 for
which the fluence indicated an isotropic equivalent energy in
excess of $M_{\circ}c^2$  \cite{Ga99}, require large
focusing factors if the central engine is indeed associated with a
stellar mass object as widely believed, and ii) the achromatic
break seen in the light curve of afterglow emission in GRB 990510
\cite{Ha99}, can be interpreted (e.g., refs \cite{Rh97,Mo99}),
as due to a transition whereby the blast
wave decelerates from $\Gamma >1/\psi$ to $ \Gamma <1/\psi$, where
$\psi$ is the opening angle of the outflow.  A blast wave model 
yielded an opening angle of $\psi= 0.08$ and a corresponding beaming 
factor of 300 for this source \cite{Ha99}.

If GRB's are indeed collimated, as they appear to be,  this would
not be surprising.  Collimation is a generic feature of outflows
in astrophysics, and appears in outflows from protostars, AGN, and
accretion disks around neutron stars (e.g. SS433).  There is as
yet no universally accepted explanation for such collimation. One
possibility is collimation by magnetic  hoop stress that arises
when the field is wound up by rotation. Such a field
configuration, however, would be kink unstable \cite{Ei93},
and in any case is very slowly collimated \cite{Be98}.
Another possibility is simply that the inertial forces in the
outer part of the flow collimate material that impacts upon it
from within. This hypothesis merely makes the plausible assumption
that  the flow is unsteady or multicomponent, so that there are
glancing collisions between fluid elements on different
streamlines. Even a steady, one component flow can display a
collimated jet if  it originates from a ring, with some of it
directed towards the ring's axis of cylindrical symmetry \cite{PE96}.  
If that part of the fluid that converges at the
axis makes a dissipative "sticky" collision, then it remains well
collimated as seen by the external observer.

In GRB's, there is an {\it a priori} reason to suspect
multicomponent flows:  If the central engine is a black hole,
then there needs to be an accretion disk feeding it, and the
luminosity from the accretion disk would be so super-Eddington
that a baryon outflow would be dragged out with the photons and
electron-positron pairs \cite{LE93,Wo93}. This would obscure the 
gamma ray burst,
so we probably have to say that the gamma ray burst is seen as
such only when we look down the axis of the baryon poor jet (BPJ)
that emerges along field lines connected to the event horizon,
which are protected by the horizon from baryon contamination
\cite{LE93}. The energy that powers the BPJ is
either extracted from the rotation of the black hole, or deposited
there by neutral particles that have crossed over from field lines
(such as neutrinos that annihilate on the  horizon-threading field
lines).

{\it The structure of colliding outflows} - 
We consider an accelerating, baryon poor outflow (BPJ) confined by the
pressure and inertia of a baryon rich wind.  We envision that the
baryonic wind is expelled (e.g., from an extended disk or torus) over a
range of scales much larger than the characteristic size of the central
engine ejecting the BPJ, so that not too far out its streamlines may
diverge more slowly than the BPJ streamlines, thereby giving rise to a
collision of the two outflows.
In general, the collision of the two fluids will lead to
the formation of a contact discontinuity across which the total pressure
is continuous, and two oblique shocks, one in each fluid, across which
the streamlines of the colliding (unshocked) fluids are deflected.
The details
of this structure will depend, quite generally, on the parameters of
the two outflows and on the boundary conditions.

We seek to determine the structure of the shocked layers assuming that
the parameters of the unshocked fluids are given.  The problem is
then characterized by 11 independent variables: the number density,
energy density, pressure, and velocity of each of the two shocked layers,
and the
cross-sectional radii of the two shocks and the contact discontinuity
surface.  In what follows subscripts $j$ and $b$ refer to quantities
in the BPJ and in the baryonic wind, respectively, and subscript $s$
denotes quantities in the shocked layers.  To simplify the analysis we
approximate the shocked layers as cylindrically symmetric, coaxial, one
dimensional flows along the channels, and denote by $a_c$,
$a_j$, and $a_b$, respectively, the cross sectional radii of the contact
discontinuity surface, the inner shock surface (in the BPJ), and the outer
shock surface.  The energy flux incident into the shocked BPJ
layer through the shock is given by
\begin{equation}
T^{0k}_{j}n_k=(p_j+\rho_jc^2)\Gamma_j U_j\sin\delta_j,\\
\end{equation}
and the momentum flux in the axial direction incident through the shock is
\begin{equation}
T^{zl}_{j}n_l=p_j+(p_j+\rho_jc^2)U_j^2\cos\theta_j\sin\delta_j.
\end{equation}
Here $n_k$ are the components of a unit vector normal to the shock surface,
$\theta_j$ is the angle between the velocity of the fluid just upstream
the shock and the BPJ axis, $\Gamma_j$ and $U_j$ are the Lorentz factor
and 4-velocity of the unshocked BPJ,
and $\delta_j$ is the deflection angle (i.e., the angle between the velocity
of upstream fluid and the shock surface), which is related to $\theta_j$
through: $\sin\delta_j= (\sin\theta_j+\cos\theta_j da_j/dz)
/\sqrt{1+(da_j/dz)^2}$.  With the above results
the energy and momentum equations of the shocked fluid can be written as
\begin{equation}
\frac{d}{dz}[T^{0z}_{sj}\pi(a_c^2-a_j^2)]=T^{0k}_{j}n_k dS/dz=
(p_j+\rho_jc^2)\Gamma_j U_j\pi a_j(\sin\theta_j+\cos\theta_j da_j/dz),
\end{equation}
and
\begin{equation}
\frac{d}{dz}[T^{zz}_{sj}\pi(a_c^2-a_j^2)]=T^{zl}_{j}n_ldS/dz=
p_j+(p_j+\rho_jc^2)U_j^2\cos\theta_j
\pi a_j(\sin\theta_j+\cos\theta_j da_j/dz),
\end{equation}
where $T^{0z}_{sj}$, $T^{zz}_{sj}$ are the energy and momentum
fluxes in the axial (z) direction of the shocked BPJ fluid, and
$dS=\pi a_j\sqrt{1+(da_j/dz)^2}dz$ is
an element of the shock surface between $z$ and $z+dz$.
The projection of the momentum equation on the
direction normal to the shock surface yields
\begin{equation}
p_{sj}=T_j^{kl}n_{j}n_k=p_j+(p_j+\rho_jc^2)U_j^2\sin^2\delta_j.
\end{equation}
The last equation holds provided the transverse momentum flux incident
through the shock exceeds the free expansion pressure of the shocked
fluid.  The above equations must be supplemented by an equation of state,
$p_{sj}=p_{sj}(\rho_{sj})$, and a continuity equation that governs the
density profile of the shocked fluid.  Likewise, the equations describing
the structure of
the shocked baryonic wind can be written in the same form with the subscript
$j$ replaced by $b$.  Finally, the requirement that the net momentum flux
across the contact discontinuity surface vanishes implies that
$p_{sb}=p_{sj}$.

A complete treatment requires numerical integration of the coupled set of
11 equations derived above, subject to appropriate boundary conditions,
and is left for future investigation.  Below, we consider for illustration
a simple situation where the BPJ is assumed to be pressure dominated, and the
assumption is then justified a posteriori.  This
corresponds
to the limit $a_j=\sin\delta_j=0$.  We further suppose that the shocked
baryonic layer is very thin, so that to a good approximation we can set
$a_b=a_c$ in the above equations.  Adopting a relativistic equation of state
for the BPJ, $p = \rho c^2/3$, mass, energy and momentum conservation imply:
\begin{eqnarray}
\label{eq:con}
p^{3/4}_{sj}\Gamma_{sj} \pi a_c^2 =C,\\
\label{eq:enrg}
(p_{sj}+\rho_{sj} c^2)\Gamma_{sj}^2\pi a_c^2c \simeq 4p_{sj}\Gamma_{sj}^2
\pi a^2_cc=L_{j},\\
\label{eq:mom}
p_{sj}=p_b+(p_b+\rho_bc^2)U^2_b\sin^2\delta_b,
\end{eqnarray}
where C and $L_j$, the corresponding BPJ power, are constants.
Equations (\ref{eq:con}), (\ref{eq:enrg}) can be combined to yield a
relation between the Lorentz factor, the pressure, and the cross sectional
radius:

\begin{equation}
(\Gamma_{sj}/\Gamma_{sjo})=(a_c/a_{co})=(p_{sj}/p_{sjo})^{-1/4},
\label{eq:Gamma}
\end{equation}
where $\Gamma_{sjo}$, $a_{co}$, and $p_{sjo}$ are the corresponding
values at $z=z_o$.  On substituting eq. (\ref{eq:Gamma}) into eq.
(\ref{eq:mom}), one obtains a differential
equation for $a_c(z)$:
\begin{equation}
(a_c/a_{co})^{-4}=p_b/p_{sjo}+\frac{(p_b+\rho_bc^2)U_b^2}{p_{sjo}}
\frac{(\sin\theta_b+\cos\theta_b da_c/dz)^2}{[1+(da_c/dz)^2]}.
\label{eq:profile}
\end{equation}

{\it Pressure confinement} - 
We consider first the possibility that the BPJ is collimated by the
pressure of the baryonic wind.  Such collimation will occur naturally
in the region located within the acceleration zone of the BPJ and above the
acceleration zone of the baryonic wind (roughly above its critical point),
 even if the two outflows have the same equation of state.  The reason
is that in this region the
density profile of the BPJ declines with radius as $r^{-3}$ (for a conical
BPJ) while that of the wind declines as $r^{-2}$.  Taking $\delta_b=0$ in
equation (\ref{eq:mom}) and using eq. (\ref{eq:Gamma}) one finds,
\begin{equation}
a_c(z)=a_{co}[p_b(z)/p_{bo}]^{-1/4}.
\end{equation}
Adopting an equation of state of the form $p_b(n_b)\propto n_b^{\alpha}$
for the baryonic fluid yields, $a_c(z)=a_{co}[n_b(z)/n_{ob}]^{-\alpha/4}$.
In case of a conical (or spherical) wind, $n_b(z)\propto z^{-2}$,
and the cross sectional radius scales as $a_c(z)\propto z^{\alpha/2}$.  
Convergence occurs for $\alpha<2$, which is the case for both
relativistic gas ($\alpha=4/3$) and non-relativistic gas ($\alpha=5/3$).

The above analysis does not take into account effects associated with the
formation of a rarefaction wave in the baryonic outflow due to
the divergence of its stream lines near the interface separating the two
fluids.  This will result in a steeper decline of the
pressure supporting the BPJ at radii where the wind is highly supersonic,
and the consequent alteration of the profile of the
contact discontinuity surface.  We naively expect collimation to take
place predominantly over a range of radii at which the Mach number of
the baryonic wind is mild, unless some fraction of the wind power is
dissipated, e.g., due to formation of shocks, far enough out.  This
suggests that only a modest beaming factor may be attainable in this case.
The determination of the level of collimation in this scenario is left
for future work.

{\it BPJ confinement by a wind emanating from a torus} - 
As a second example we consider a baryonic wind emanating from
a thin torus of radius $R$ located at $z=0$, where z is the axial
coordinate, and centered around the central engine ejecting the
BPJ (see Fig. 1).  We suppose that on scales of interest the
baryonic wind is highly supersonic, so that momentum transfer from
the wind into the BPJ is dominated by ram pressure. We can therefore
neglect the kinetic pressure, $p_b$, in eq. (\ref{eq:profile}).
Now the ram pressure of the baryonic outflow just upstream the oblique
shock at some position $z$ is related to the total wind power
, $L_b$, through:
\begin{equation}
(p_b+\rho_bc^2)U_b^2=\beta_bL_b/(4\pi^2 c a_c r),
\label{eq:Lb-def}
\end{equation}
where $r=[z^2+(R-a_c)^2]^{1/2}$ is the distance between a
point on the torus and the nearest point to it at the shock (see figure 1),
and $\beta_b$ is the
(terminal) wind velocity.  The angle between the wind velocity and
the BPJ axis is given by $\tan\theta_b=(R-a_c)/z$.  Substituting eq.
(\ref{eq:Lb-def}) into eq. (\ref{eq:profile}), and using
eq. (\ref{eq:enrg}) yields,
\begin{equation}
\frac{(R-a_c+da_c/d\ln z)^2}
{1+(da_c/dz)^2}=a_{co}^2\chi^{-1}\frac{[z^2+(R-a_c)^2]^{3/2}}{a_c^3}
\label{eq:tor}
\end{equation}
with $\chi=\pi^{-1}\beta_b \Gamma^2_{sjo}(L_b/L_j)$ being approximately
the ratio
of baryonic wind and BPJ powers.  At large enough axial distances from the
plane of the torus, $z>>R$, an approximate solution to eq. (\ref{eq:tor})
can be obtained by expanding the various terms in powers of $R/z$, and
becomes more exact at large z. To
second order the BPJ in this regime is conical, viz., $a_c(z)=b+\alpha z$,
where $\alpha$, the opening angle, and $b$ are constants, and satisfy the
relation:
\begin{equation}
\chi(R/a_{co}-b/a_{co})^2=\frac{(1+\alpha^2)^{5/2}}{\alpha^3}.
\end{equation}
The parameter $b/a_{co}$ depends on the properties of the solution
near the central engine (at $z<R$) and must be determined numerically.
The requirement that the BPJ is confined by the ram pressure of the
baryonic outflow at $z=0$ yields,
\begin{equation}
(R/a_{co}-1)=\chi,
\end{equation}
where it has been assumed that $da_c(z=0)/dz=0$.
Combining the last two equations, one finds
\begin{equation}
\frac{(1+\alpha^2)^{5/2}}{\alpha^3}=\chi(\chi +1-b/a_{co})^2.
\label{eq:ang}
\end{equation}
For $\chi>>1$ the latter equation implies that
\begin{equation}
 \alpha\simeq
\chi^{-1}+O(\chi^{-2}).
\end{equation}
Numerical integration of eq. (\ref{eq:tor}) confirmed that the BPJ
is indeed conical with an opening angle given by eq. (\ref{eq:ang}),
except for a small range ($z<R$) near the central engine.
BPJ profiles computed numerically for different values of the parameter
$\chi$ are exhibited in Fig. 2, and the corresponding dependence of
$\alpha$ on $\chi$ is displayed in Fig. 3.

Once the BPJ Lorentz factor exceeds $\alpha^{-1}\sim \chi$, it will remain
conical with the same opening angle regardless of the external conditions.
This occurs at a distance $z=R\chi^2/(1+\chi)$ from the central engine.
Thus the baryonic wind must extend to at least this scale.

The apparent luminosity, $L_{app}$, exhibited by a beamed source
having a power $L_j$ is roughly $L_{app}\sim f^{-1}L_j$, where
$f=\Delta \Omega/2\pi$ is the corresponding beaming factor.  For the
above model we find
\begin{equation}
f=\pi \alpha^2/2\pi=1/2\chi^2=(\pi^2/2)(L_j/L_b)^2,
\label{eq:beaming}
\end{equation}
and
\begin{equation}
L_{app}\sim (2/\pi^2)(L_b/L_j)L_b.
\label{eq:Lapp}
\end{equation}


    In conclusion, we have shown [equation (\ref{eq:Lapp})] that an inner 
relativistic jet collimated by an outer
modestly relativistic jet can appear brighter to the observer within the beam
by a factor that is {\it inversely } proportional to the intrinsic power of the
inner jet. When all else is the same, a  low luminosity inner jet actually
appears {\it brighter} than a high luminosity one  to
the observer within its beam. The cost, of course, is that, fewer observers
are within the beam.  Ironically, most of the energy of the GRB is the
 {\it unseen} outer jet.

    Now consider the remarkable GRB 990123, which had an isotropic equivalent
energy output of about $3 \times 10^{54}$ ergs.  Applying equations
(\ref{eq:beaming}) through (\ref{eq:Lapp}), we propose that the actual energy
in the gamma ray fireball  was only about $\eta ^2 10^{50}$ ergs, (it may
have even been
{\it smaller} than average) and an outer baryonic wind output of order
$ \eta 10^{52.5}$ erg. The corresponding beaming factor is
$f=10^{-5} \pi^2\eta^2/2$. Here $\eta $ is assumed to be of order 
unity or less, otherwise the energetics strain theoretical models.

A rough  observational lower limit on $\eta$ can in principle be derived
from the consideration that GRB 990123 could not have a collimation
angle that is implausibly smaller than for
average GRB's.  Although it was an unusual burst, the {\it a priori} chance
of seeing it is unlikely to have been less than $10^{-3}$, and, considering
that it is a BeppoSax-located GRB and that there has been at least one
other burst of comparable (if somewhat smaller) intrinsic magnitude,
probably considerably more than $10^{-3}$.
The largest uncertainty is the beaming factor for the {\it average} GRB.
If we take that as being of order $10^{-3}$, corresponding to an opening
angle of about 0.06 radians, this suggests a value for $\eta$ of at least
of order 0.3.  This means that the minimum energy requirement for GRB
990123 is still of order $10^{52}$ ergs or more.  On the other hand, the fact
that it can come out as a baryonic wind relaxes the demands on theoretical
models.  For example, a magnetocentrifugal wind could plausibly release 
several percent of a solar rest energy over 30 seconds, whereas it would 
be hard to get this much from $\nu - \bar\nu$ annihilation. 

Another example is GRB990510, for which the achromatic steepening of the 
optical and radio light curves observed after about 1 day \cite{Ha99}, 
can be modeled as due to the evolution of a jet with an opening angle 
of $\sim 0.08$ rad, and a corresponding beaming factor of 300.  The 
prompt GRB emission should be at least as beamed as the afterglow emission,
but could in fact be more beamed.
The isotropic gamma-ray emission inferred for this 
source is $3\times10^{53}$ ergs.  Employing equations (\ref{eq:beaming}) 
and (\ref{eq:Lapp}) yields a total energy of $<10^{51}$ ergs for
the BPJ and $3\times 10^{52}$ ergs for the confining wind. 

We acknowledge support from the Israel Science Foundation

\newpage
-
\begin{figure}
\vspace{15cm}  
\includegraphics{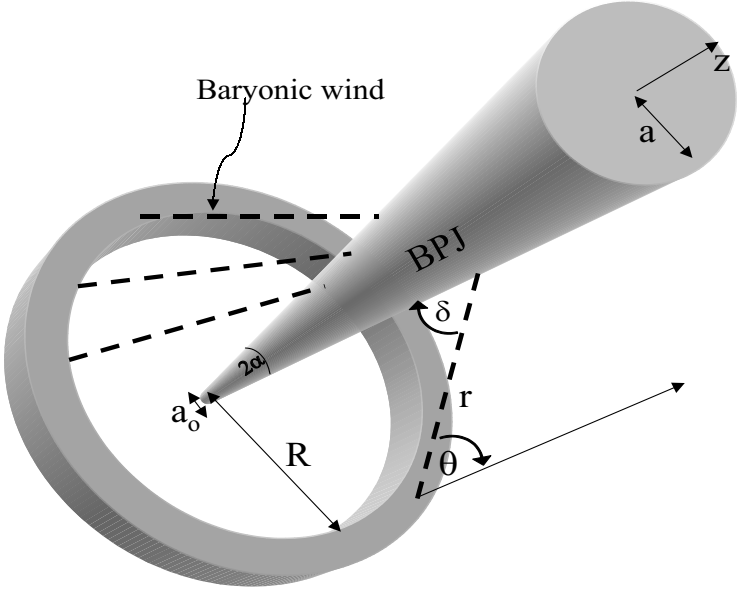}
\caption{Schematic illustration of BPJ confinement by the ram
pressure of a baryon rich wind (represented by the dashed lines)
emanating from a torus of radius $R$.  The torus is centered
around the central engine ejecting the BPJ.  Fluid elements of
the baryonic outflow collide with the BPJ at an angle $\delta(z)$,
and are deflected when passing through an oblique shock.
The shocked wind layer is assumed to be very thin.
At distances larger than roughly $R$ the BPJ becomes
conical with a semi-opening angle $\alpha$.}
\end{figure}        

\newpage
-
\begin{figure}
\vspace{12cm}  
\includegraphics{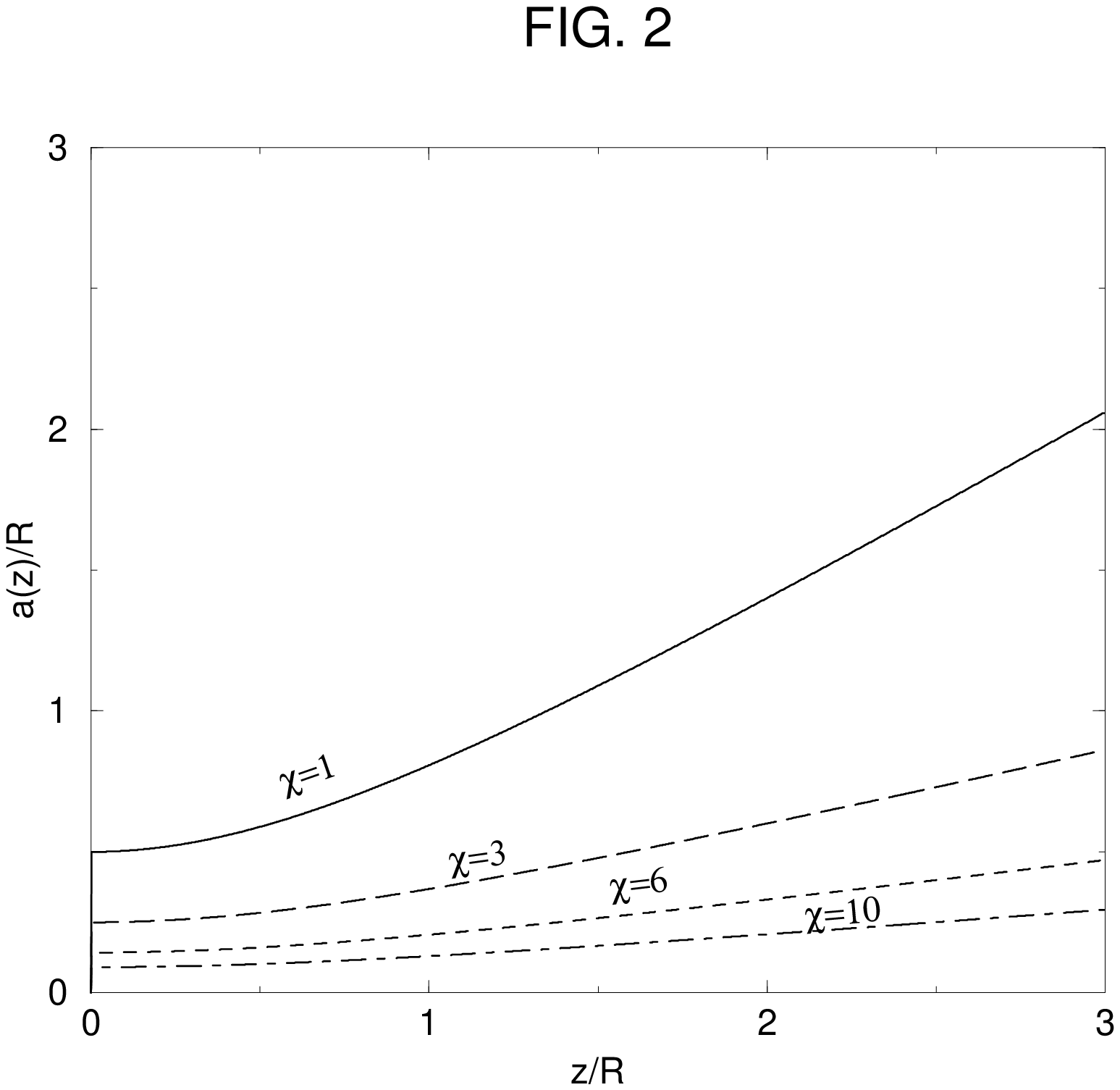}
\caption{Cross sectional radius of the BPJ versus axial
distance from the central engine, computed numerically
for different values of $\chi$.}
\end{figure}
\newpage
-  
\begin{figure}
\vspace{12cm}  
\includegraphics{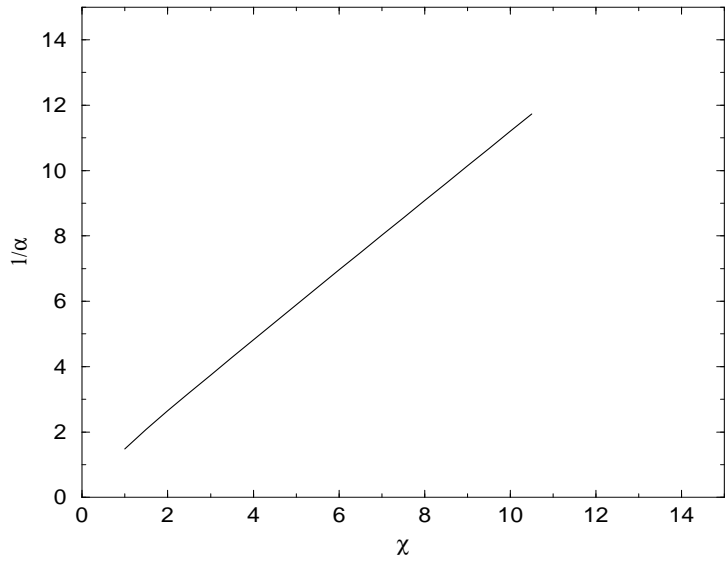}
\caption{A plot of $\alpha^{-1}$against $\chi$, obtained from
the numerical integration of eq. (\ref{eq:tor})}
\end{figure}
\end{document}